\documentclass[runningheads]{llncs}
\UseRawInputEncoding
\usepackage{graphicx}
\usepackage{algorithm,algorithmic}
\begin{document}
\title{Working mechanism of Eternalblue and its application in ransomworm}
%
%\titlerunning{Abbreviated paper title}
% If the paper title is too long for the running head, you can set
% an abbreviated paper title here

 \author{Zian Liu 
 }
% %
 \authorrunning{Z. Liu et al.}
% % First names are abbreviated in the running head.
% % If there are more than two authors, 'et al.' is used.
% %
\institute{Swinburne University of Technology, Victoria 3122, Australia
%\email{lncs@springer.com} 
 }
\maketitle              % typeset the header of the contribution
\begin{abstract}
After the leaking of exploit Eternalblue, some ransomworms utilizing this exploit have been developed to sweep over the world in recent years. Ransomworm is a global growing threat as it blocks users' access to their files unless a ransom is paid by victims. Wannacry and Notpetya are two of those ransomworms which are responsible for the loss of millions of dollar, from crippling U.K. national systems to shutting down a Honda Motor Company in Japan. Many dynamic analytic papers on Wannacry were published, however, static analytic papers about Wannacry were limited. Our aim is to present readers an systematic knowledge about exploit Eternalblue, from a high\textendash leveled semantic view to the code details. Specifically, the working mechanism of Eternalblue, the reverse engineering analysis of Eternalblue in Wannacry, and the comparison with the Metasploit's Eternalblue exploit are presented. The key finding of our analysis is that the code remains almost the same when  Eternalblue is transplanted into Wannacry, which indicates its potential for signatures and thus detection. 

\end{abstract}
\section{Introduction}
With computers and networks being applied more and more widely in daily life, enterprise and organizations tend to store large scale data digitally. However, those information management systems often contain vulnerabilities and are prone to be exploited by hackers. 
%Healthcare organizations are one of the biggest victims in this kind of attack. According to \cite{c21}, the average attack the other organizations suffer is 14,300 per day, while in 2017, average healthcare organizations suffer 32,000 intrusion attacks per day. Furthermore, personal health data is 50 times more valuable than financial information on black market. The stolen patient data can worth \$50 per record.
In May 2017, a ransomworm called \textit{Wannacry} bursts out worldwide. It caused massive infection and enormous economic losses by infecting various industrial and government internal networks, such as UK's National Health Service and etc \cite{c16}. The average attack the other organizations suffer is 14,300 per day according to \cite{c21}. It is reported to derive from an NSA exploit tool \cite{c8}. Once Wannacry infects a system, it encrypts copies of various file types and deletes the originals. The encrypted files cannot be accessed without a decryption key.

A lot of work has been done to analyze Eternalblue and Wannacry. Most of the papers on Wannacry focused on the dynamical analysis. D.Y. KAO et al. analyzed Wannacry dynamically, from the aspects of process name, Registry, file system, and Network activity, respectively. They also applied the features to these aspects to create Yarra rules for pattern-matching detection \cite{c14}. Qian et al. also dynamically analyzed Wannacry for testing the performance of an automatic dynamical analysis tool \cite{c18}. As to the static analysis, D.Y. KAO et al. provided a detailed analysis based on different phases. Critical files and strings participating in those phases were highlighted \cite{c19}. Hirokazu statically analyzed Wannacry based on the Eternalblue and DoublePulsar modules. He also applied the discoveries into Snort rules to defend future network attacks based on those two modules \cite{c20}.

Even though those works helped investigate the working mechanism of Wannacry and Eternalblue, there lacks the study on the comparison between Wannacry's Eternalblue module and the original Eternalblue module. To bridge this research gap, we proceed with our analysis by first studying the exploit's working mechanism that applied in \textit{Wannacry}. Then we use code analytic tools and network capture tools to compare Wannacry's Eternalblue module and the original Eternalblue module. After a detailed study, we find that the exploit utilized in Wannacry shares a very similar pattern to the original exploit, which can be used as features for signature extraction.

The remainder of this paper is organized as follows. In section 2, the original Eternalblue module's working mechanism is introduced. Section 3 analyzes the Eternlablue module in \textit{Wannacry} and Section 4 concludes this paper.

\section{Eternalblue's working mechanism in Metasploit}
Assume we have two computers: an attacking machine and a victim machine. At the very beginning, Eternalblue is a piece of code on the attacking machine. Once executed, it will send multiple SMB (Server Message Block) requests to the victim machine through the SMB protocol. As a result, the victim machine must respond to these requests. In this SMB communication phase, the attacking machine plays the role of client and the victim machine plays the role of server, which is the reason we refer them to attacking machine and victim machine respectively. Among these SMB requests, the Transaction SMB commands are essential because they are utilized to tamper the data on the server (victim) with a buffer overflow bug, which further leads to the execution of the ransomworm on the victim machine.

Hence, prior to introducing details of the working mechanism of Eternalblue, in this section we will firstly introduce the normal usage of Transaction SMB. According to MSDN \cite{c10}, Transaction SMB commands enable the client to access advanced features on the server. Specifically, the three transaction messages are:
\begin{itemize}
    \item SMB\_COM\_TRANSACTION (or \textit{Trans}),
    \item SMB\_COM\_TRANSACTION2 (or \textit{Trans2}),
    \item SMB\_COM\_NT\_TRANSACT (or \textit{NT Trans}).
\end{itemize}

It is also noted in \cite{c11} that, SMB\_COM\_NT\_TRANSACT subcommands enable the transfer of very large data chunks. And SMB\_COM\_TRANSACTION2 subcommands provide richer file system services such as allowing clients to set and retrieve Extended Attribute key/value pairs, to make use of long file names, and to perform directory searches, etc.  

The information above summarizes the legitimate usage of the Transaction SMB commands. However, in Eternalblue, they are not applied for the original legitimate purposes, but for a buffer overflow bug. Specifically, when responding to these crafted requests, the server will convert the payload contained in these requesting Transaction SMB packets, i.e., the original Os2Fea \cite{c5} list, to the currently used NtFea \cite{c5} format (result type), as Os2Fea (the original type) is outdated \cite{c12}. The NtFea list will be stored into a \textit{result list buffer}. This process is also referred to the conversion process. Under some mild conditions, the server can be fooled by allocating a result list buffer smaller than the NtFea list to be stored. Thus the NtFea list can overwrite the next buffer. And the original list with a specific length will satisfy such mild conditions.

In more detail, the ``next buffer" is a Srvnet.sys \cite{c5} buffer, which is allocated on the server for the attacker's SMB request. Once allocated, this buffer will wait for another data package to be sent to the server. There are two parameters in the header of this Srvnet.sys buffer: one decides where to map the data package on this server upon receiving the data package and another decides what function to execute when the Srvnet connection is disconnected. So if these two parameters are modified to the same address, the payload will be mapped and be executed upon closing the connection.

To trigger the overflow of Srvnet.sys and thus inject the malicious codes to the victim, Eternalblue covers three essential steps: \textit{crafting original list}, \textit{buffer grooming}, and \textit{sending the payload}. We proceed our discussion of Eternalblue by first showing a high-level description of the three steps, then dive into details of each step.  

From a high-level point of view, in step \textit{crafting original list}, an original list is crafted. In the second step, multiple grooming packages are sent in a deliberate order which changes the server's buffer status to a point that is vulnerable to overflow. Then, sending the complete original list results in the overflow. In the final step, the payload is sent to the server's Srvnet.sys buffer. Because of the overflow, the payload can be mapped to the desired location and executed upon closing the connection.

\subsection{Crafting original list}
To understand crafting of the original list, we firstly recall the normal conversion process on the server machine. 

As shown in Algorithm \ref{O2Fea}, in the Os2Fea format, there is a parameter \textit{ULONG SizeOfList} prior to the actual records describing the total bytes of the original list.

\begin{figure}[ht]
 \centering
      \includegraphics[scale=0.3]{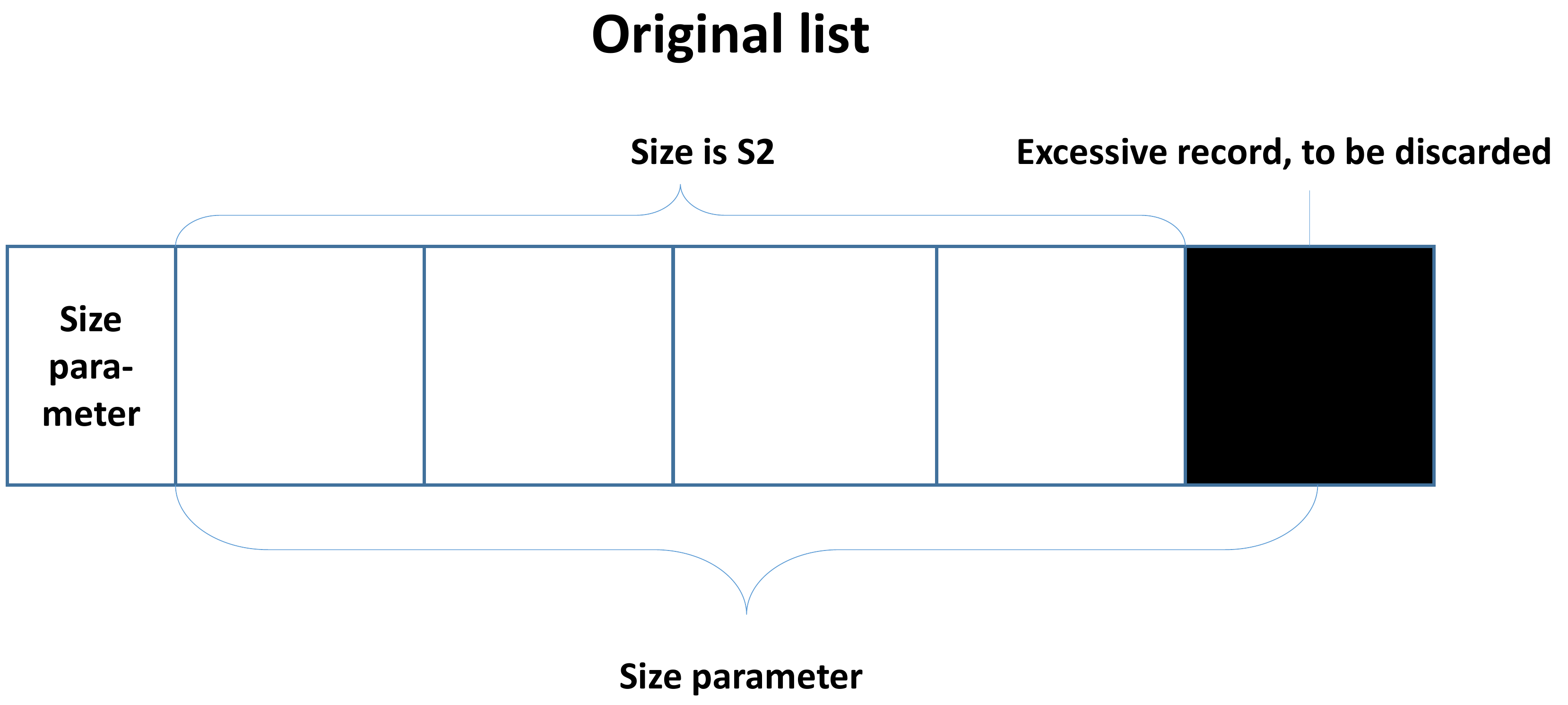}
    
      \caption{Server discards the out-of-boundary records and calculates the result list size (S1) based on the remaining records}\label{discard}
     
   \end{figure}

The server's legitimate conversion process is shown in Algorithm \ref{algorithm1}. In step \textit{Compute S1}, the algorithm will go through the original list and discard the records that exceed the boundary set by SizeOfList, and the remaining original list (Os2Fea) size is S2, as shown in Fig. \ref{discard}. S1 is the result list (NtFea) size corresponding to the remaining original list with size S2. Hence, in the third line of Algorithm \ref{algorithm1}, the server allocates the result list buffer with size S1. 

Back into the second line of Algorithm \ref{algorithm1}, S2 needs to be assigned to the original list's parameter SizeOfList, as this parameter will be used later in the while loop. In the \textit{while} statement of Algorithm \ref{algorithm1}, the server calls a subfunction repeatedly to convert the original list block by block and stores the result list into this result list buffer. The number of the loop is determined by the SizeOfList's value. The original\_list\_initial\_address in this algorithm points to the beginning of the original list. It is later assigned to variable \textit{Current\_pointer}, which will increase after each iteration.

Above is the server's conversion process. The bug occurs when assigning S2 to the original list's parameter SizeOfList if the  SizeOfList is no less than $2^{16}$ (0x10000 in hexadecimal) and the actual original list's record exceeds the boundary set by SizeOfList \cite{c2}, like shown in Fig. \ref{discard}. 

In more detail, when parsing S2 to SizeOfList, only the LOWORD (low-order word) bytes  of the DWORD (double word) variable SizeOfList is updated because of a wrong casting instruction. Hence the End\_pointer in Algorithm \ref{algorithm1} will be miscalculated, which leads to an unexpected conversion time. This corresponds to a different time to execute the \textit{while} statement in Algorithm \ref{algorithm1}. For example, as occurred in this Eternalblue exploit, the SizeOfList is initiated with 0x10000. After discarding, the remaining original list size S2 is 0xff5d. However, when executing the second line of Algorithm \ref{algorithm1} (i.e., assigning S2 to SizeOfList), only the LOWORD bytes of SizeOfList is updated, which turns SizeOfList from 0x10000 to 0x1ff5d rather than 0xff5d, hence enlarge this SizeOfList.

   \begin{algorithm}[ht]
\caption{Os2Fea list structure}
\label{O2Fea}
 \begin{algorithmic}
 \STATE $Struct\ Os2FeaList\{$
 \STATE $ULONG\ SizeOfList$
 \STATE $UCHAR\ Os2FeaList\lceil SizeOfList-4\rceil$
 \STATE $\}$
 \end{algorithmic}
 \end{algorithm}
 
\begin{algorithm}[ht]
\caption{The legitimate server's conversion process}
\label{algorithm1}
 \begin{algorithmic}
 \STATE $Compute\  S1$
 \STATE $S2\ assigned\ to\ SizeOfList\ $ 
 \STATE $Allocate \ buffer\ (result\ list\ buffer)\ with\ size\ S1$
 \STATE $End\_pointer=original\_list\_initial\_address+SizeOfList$
 \STATE $Current\_pointer=original\_list\_initial\_address$
 \WHILE{Current\_pointer\textless{}End\_pointer}
 \STATE $convert(Current\_pointer)$
 \STATE $Current\_pointer+=each\_record\_size$
 \ENDWHILE
   \end{algorithmic}
   \end{algorithm}
   
Based on the discussion above, the crafted list is as follows. The forged original list is of the Os2Fea type, its parameter SizeOfList is with value $2^{16}$ (0x10000 in hexadecimal), followed by a list of Os2Fea data, as demonstrated in Algorithm \ref{O2Fea}. There are 607 pieces of data included in this crafted list and garbage data at the end which confines the request packet to a particular size. The first 605 pieces of records are empty, the 606th record is not empty and can be filled with arbitrary data of a certain length. The 607th record contains the fake Srvnet.sys header and this 607th record exceeds the boundary set by SizeOfList \cite{c12}. As analog in Fig. \ref{discard}, the 607th record is the black section, followed by some garbage data. After discarding, only the first 606 records should be converted.

When converting this crafted list on the server, as demonstrated in Algorithm \ref{algorithm1}, after discarding, S1 and S2 are calculated representing the first 606 records of the result list and the original list. Then, the SizeOfList should be assigned to S2 but in fact assigned to an enlarged value because of the wrong casting (assigned 0x1ff5d rather than 0xff5d). Hence the \textit{End\_pointer} is also enlarged. Afterwards, the result list buffer that can only store the first 606 result records will be allocated. Later, the conversion begins, and the loop will be executed for extra times as the \textit{End\_pointer} is enlarged. Hence the server will convert and store 607 records in the buffer for 606 records. This is the reason why this 607th record shall be crafted with a forged Srvnet.sys buffer header and the preceding records can be filled with arbitrary data of a certain size. 

As mentioned earlier, once the Srvnet.sys buffer is allocated on the server, it waits for another data packet. There are two critical fields in the Srvnet.sys buffer header for processing the data packet: one is called \textit{memory descriptor list} (MDL) which points to a virtual address that the data package shall be mapped to once received； the other one is called \textit{pSrvNetWskStruct}. It points to a function which shall be called when the Srvnet connection is closed. Therefore, overwriting these two fields with the same address can make the server map the shellcode to the desired location and execute the shellcode after closing the Srvnet connection.

Finally, it should be noted that the crafted data in the 607th record should be differentiated from the shellcode since the crafted 607th record is used to overwrite the Srvnet.sys buffer's header, which paves the way for sending the shellcode. The sending process of the shellcode is to be discussed later.

\subsection{Buffer grooming}

We have discussed how the crafted list will lead to the buffer overflow in Srvnet.sys. If the buffer Srvnet.sys is not allocated exactly after the result list buffer, the attack fails. This \textit{buffer grooming} process aims to improve the success rate of overflowing the Srvnet.sys buffer. Table. \ref{table1} shows all the grooming packages sent by Eternalblue in timeline. We have validated the packets by analyzing the packets sent by the samples from 2 sources \cite{c3} and \cite{c4}. The ultimate goal of the grooming procedure is to make the server allocate a \textit{Srvnet.sys buffer} immediately following the \textit{result list buffer}. Only when this goal is achieved, the excessive data from the \textit{result list buffer} can overwrite the \textit{Srvnet.sys buffer}'s header later. The order of the packages sent in Table. \ref{table1} can increase the possibility of achieving our ultimate goal. However, the proof is complicated and out of the scope of this paper.

\begin{table}[]
\caption{Eternalblue package sent in timeline}
\label{table1}
\begin{tabular}{|l|l|l|}
\hline
No.                                                          & Type                                                   & Description                                                                                                                                                                                                                                                                                                                                                                                      \\ \hline
1                                                            & Srv                                                    & \begin{tabular}[c]{@{}l@{}}Anonymous login and IPC\$ tree connect, then send the crafted original list\\ except the last segment to the server through an NT Trans Request and\\ multiple Trans2 Secondary Requests. An Echo package is followed to ensure the \\ list was sent successfully.\end{tabular}                                                                                  \\ \hline
2                                                            & \begin{tabular}[c]{@{}l@{}}1st\\ reserve\end{tabular}  & \begin{tabular}[c]{@{}l@{}}Send malformed Negotiate Protocol Request and Session Setup AndX Request\\ to reserve buffer (0x10000 bytes) with size smaller than the result list buffer in \\NonPagedPool on the server.\end{tabular}                                                                                                                                                          \\ \hline
3-15                                                         & Srvnet                                                 & \begin{tabular}[c]{@{}l@{}}Send multiple TCP packages to establish Srvnet connections which fill up the\\ slot before the result list buffer.\end{tabular}                                                                                                                                                                                                                                     \\ \hline
16                                                           & \begin{tabular}[c]{@{}l@{}}2nd \\ reserve\end{tabular} & \begin{tabular}[c]{@{}l@{}}Send malformed Negotiate Protocol Request and Session Setup AndX Request\\ to reserve buffer (0x11000) slightely bigger than the result list buffer. This \\reserved buffer serves as a place holder for the result list buffer.\end{tabular}                                                                                                                  \\ \hline
2                                                            & \begin{tabular}[c]{@{}l@{}}1st \\ reserve\end{tabular} & \begin{tabular}[c]{@{}l@{}}Send a FIN TCP package to free the 1st reserved buffer.\end{tabular}                                                                                                                                                                                                                                                                                                \\ \hline
17-22                                                        & Srvnet                                                 & \begin{tabular}[c]{@{}l@{}}Send TCP packages to establish extra Srvnet connections. One of them is \\expected to be allocated next to the 2nd reserved buffer.\end{tabular}                                                                                                                                                                                                                \\ \hline
16                                                           & \begin{tabular}[c]{@{}l@{}}2nd \\ reserve\end{tabular} & \begin{tabular}[c]{@{}l@{}}Send a FIN TCP package to free the 2nd reserved buffer.\end{tabular}                                                                                                                                                                                                                                                                                               \\ \hline
1                                                            & Srv                                                    & \begin{tabular}[c]{@{}l@{}}Send the last segment of the original list through a Trans2 Secondary Request.\\ So the Srv.sys will convert the list. To store the result list with size 0x10fe8 \\(S1), the server allocates 0x11000 bytes. Because of Windows memory's\\ last-in-first-out working fasion, the  2nd reserved buffer just being freed should \\be allocated here.\end{tabular} \\ \hline
\begin{tabular}[c]{@{}l@{}}3-15 and \\ 17-22\end{tabular} & Srvnet                                                 & \begin{tabular}[c]{@{}l@{}}Send the shellcode through Mutiple TCP packages. The overflow ensures the\\ shellcode be mapped to a desired location. Then close the connections.\end{tabular}                                                                                                                                                                                                                             \\ \hline
\end{tabular}
\end{table}

The following paragraphs introduce the packages sent in each step listed in Table. \ref{table1}. To validate the buffer grooming process, we reproduce the spreading process in a virtual environment and check the captured network packages listed in Table. \ref{table1}. The samples are created based on scripture on the exploit-db website \cite{c3} and Metasploit Eternalblue module \cite{c4}. The baseline of the virtual environment and experiment tools are shown below:
\begin{itemize}
    \item Virtual Machine: VMWare Workstation
    \item Client (attacker) machine OS: Windows 7 x64 SP1
    \item Client (attacker) machine IP address: 10.10.10.151
    \item Server (victim) machine OS: Windows 7 x64 SP1
    \item Server (victim) machine IP address: 10.10.10.152
    \item Analysis tools: Wireshark
\end{itemize}

\begin{figure}[ht]
      \includegraphics[scale=0.4]{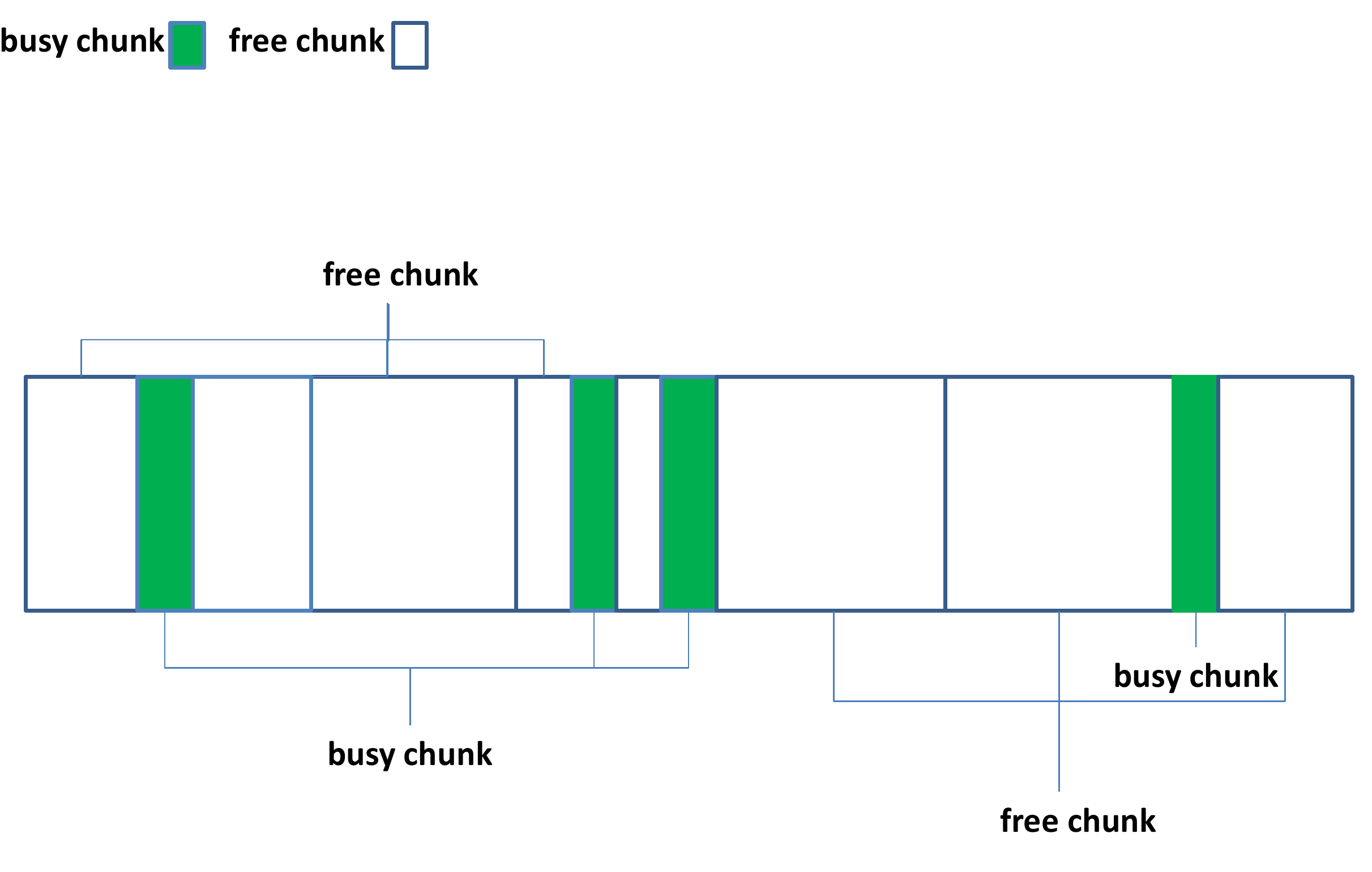}
    
      \caption{Server machine's buffer initial status}\label{buffer1}
      
   \end{figure}

Firstly the exploit from the client (attacker) machine establishes a connection and determines the target operating system's version and architecture based on the SMB and DCE/RPC (Distributed Computing Environment / Remote Procedure Calls \cite{c13}) reply, respectively. Figure. \ref{buffer1} shows the server (victim) machine's buffer initial status before receiving any packages from the client (attacker) machine.

\begin{figure}[ht]
      \includegraphics[scale=0.4]{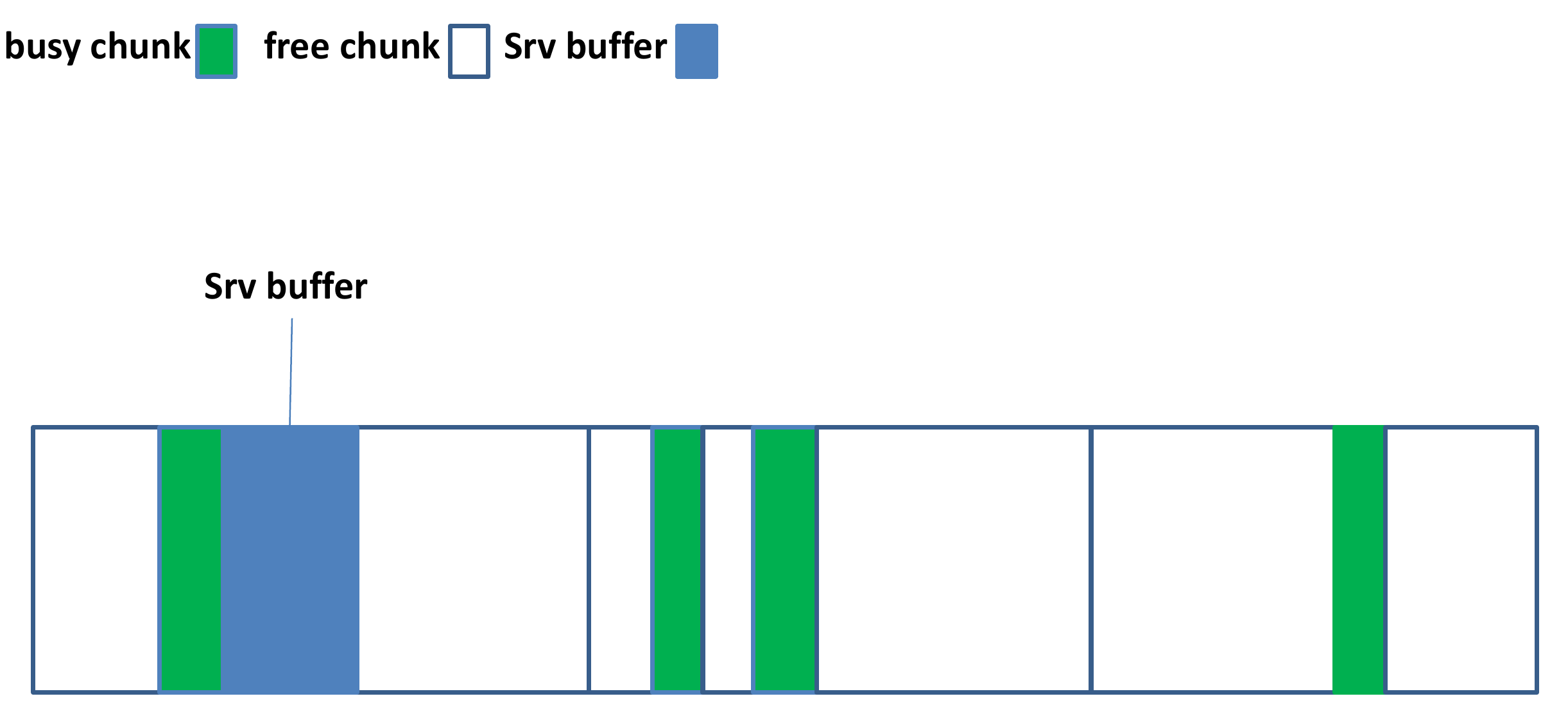}
    
      \caption{Server machine's updated buffer state}\label{buffer2}
      
   \end{figure}

Then the exploit sends the original list to the target machine through connection No.1. However, the legitimate usage of the Transaction SMB request is to send Trans2 Secondary Request after Trans2 Request or to send the NT Trans Secondary Request after the NT Trans Request. Here in this exploit, the purpose of sending Trans2 Secondary Requests packets after the initial NT Trans Request is to utilize another data parsing bug, which permits the attacker to send the payload in a Trans2 request that is bigger than its limit, e.g. 0xffff \cite{c12}. To ensure the original buffer is received correctly by the target machine, the exploit on the client machine sends an echo package to the server machine. After receiving those packages, the server machine's buffer status changes to the one depicted in Fig. \ref{buffer2}. 

\begin{figure}[ht]
      \includegraphics[scale=0.4]{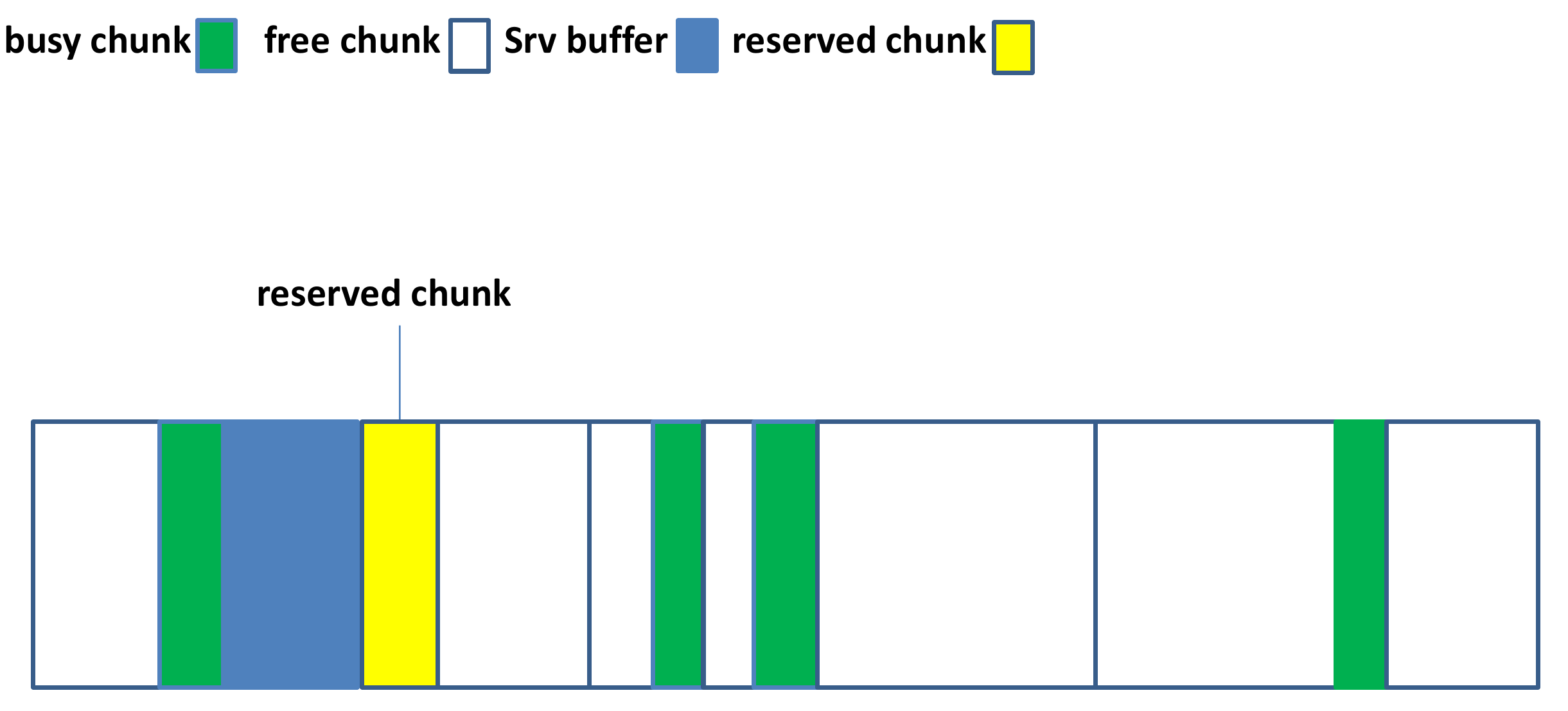}
    
  \caption{Server machine's updated buffer state}\label{buffer3}
   \end{figure}

Then in connection No.2, to reserve a buffer chunk which is used for grooming, another request is sent from the client to the server. After receiving the request, the server's buffer status is updated as shown in Fig. \ref{buffer3}.

\begin{figure}[ht]
      \includegraphics[scale=0.4]{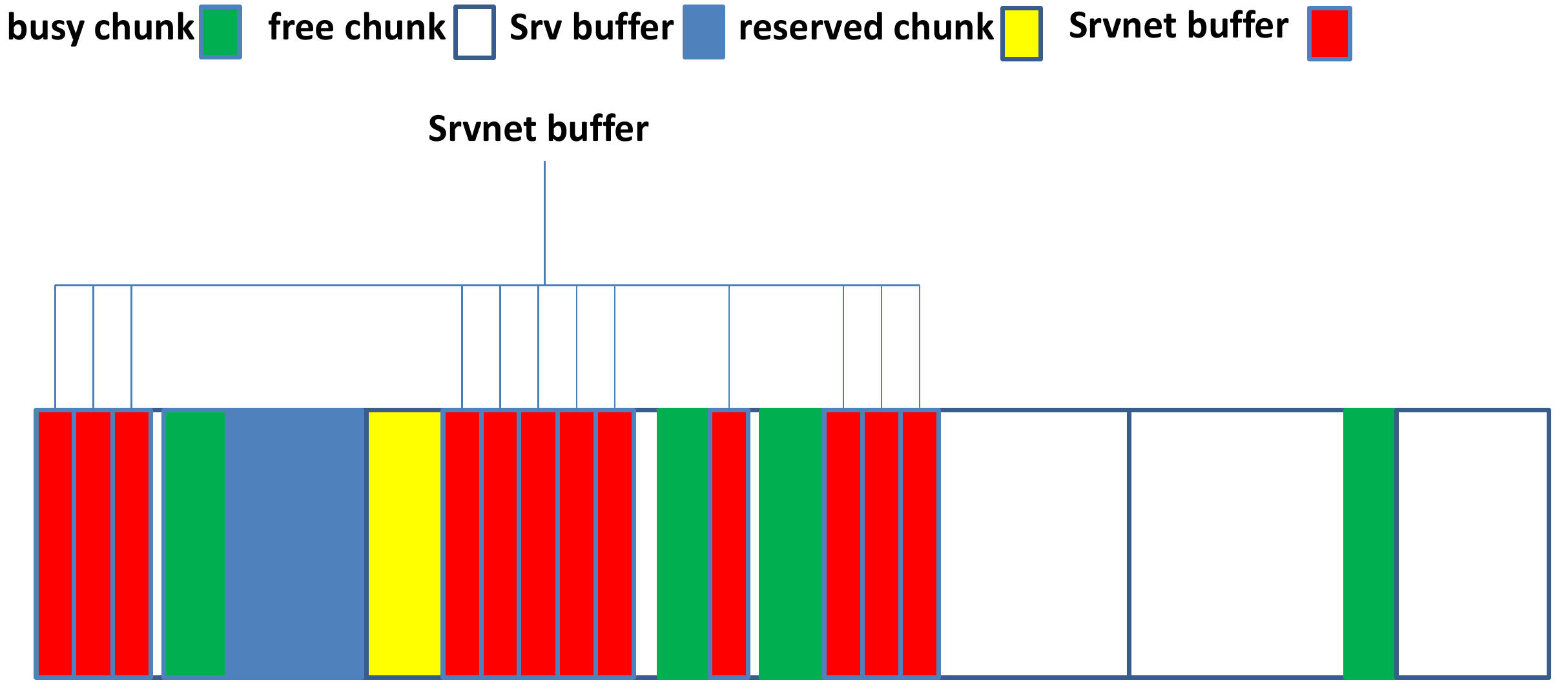}
    
      \caption{Server machine's updated buffer state}\label{buffer4}
      
   \end{figure}

Next, in order to keep grooming the buffer, as in connection No.3-15 in Table. \ref{table1}, multiple Srvnet requests are sent to allocate multiple Srvnet.sys buffer chunks on the server. This is the first series of the Srvnet request packages which fill up the slot before the second reserved buffer. Fig. \ref{buffer4} demonstrates the server's buffer status after receiving these Srvnet requests. Srvnet connections in this step increase the probability that the Srvnet buffer allocated in connections No.17-22 be allocated immediately following the \textit{result list buffer} because connections No.3-15 fill up the slot between the two reserved Srv buffer (connection 2 and connection 16).

\begin{figure}[ht]
      \includegraphics[scale=0.4]{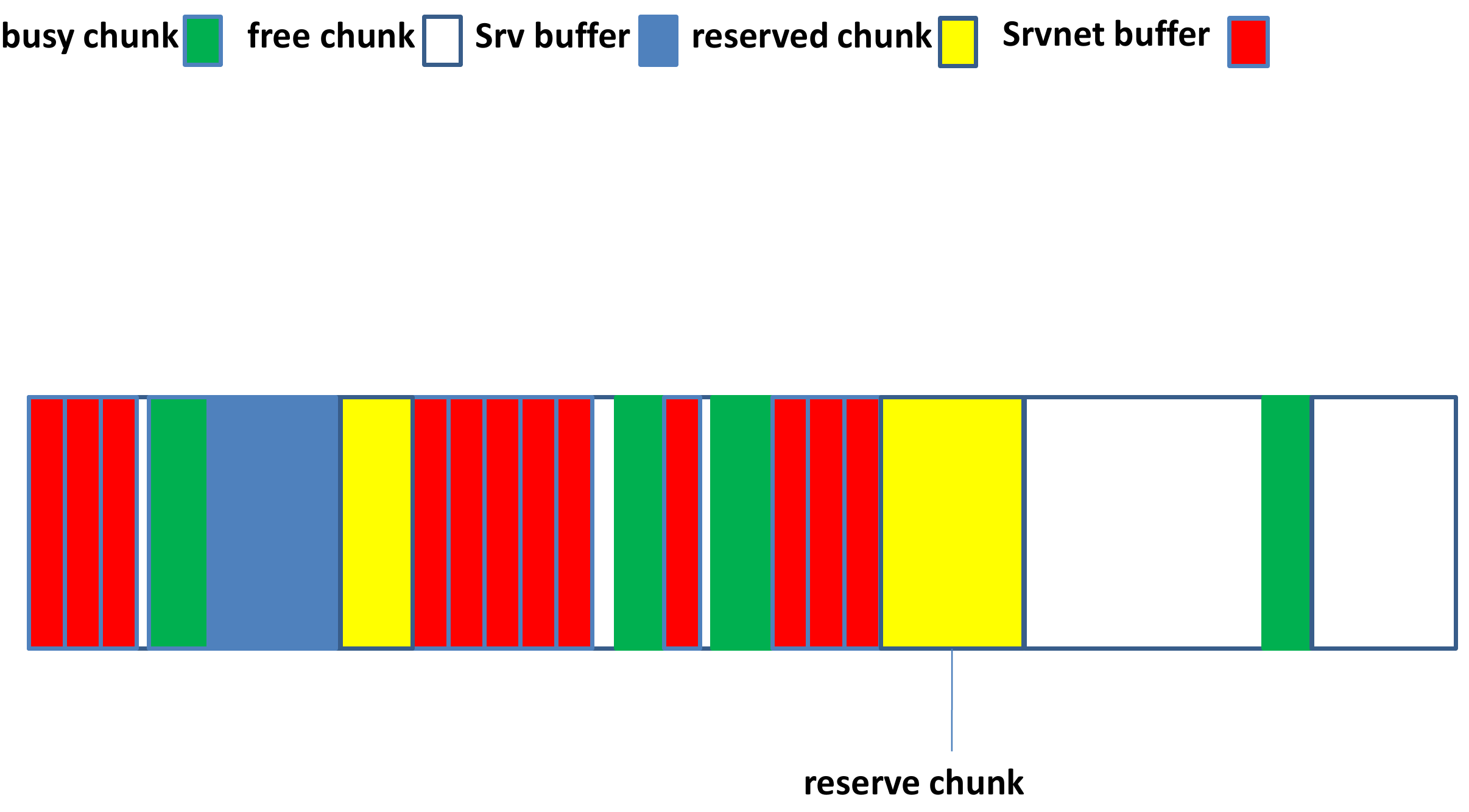}
    
      \caption{Server machine's updated buffer state}\label{buffer5}

   \end{figure}

As in connection No.16 in Table. \ref{table1}, the second reserving buffer chunk is reserved as a placeholder (to be replaced with the \textit{result list buffer} later). Afterwards, the first reserved buffer chunk through connection No.2 shall be freed. The first and second buffer reserving packets also utilize a bug by setting special parameters in the request to make the large NonPagedPool allocation \cite{c12}, which is much greater than it is permitted to. After these two steps, the server's buffer status changes to the one depicted in Fig. \ref{buffer5}.

\begin{figure}[ht]
      \includegraphics[scale=0.4]{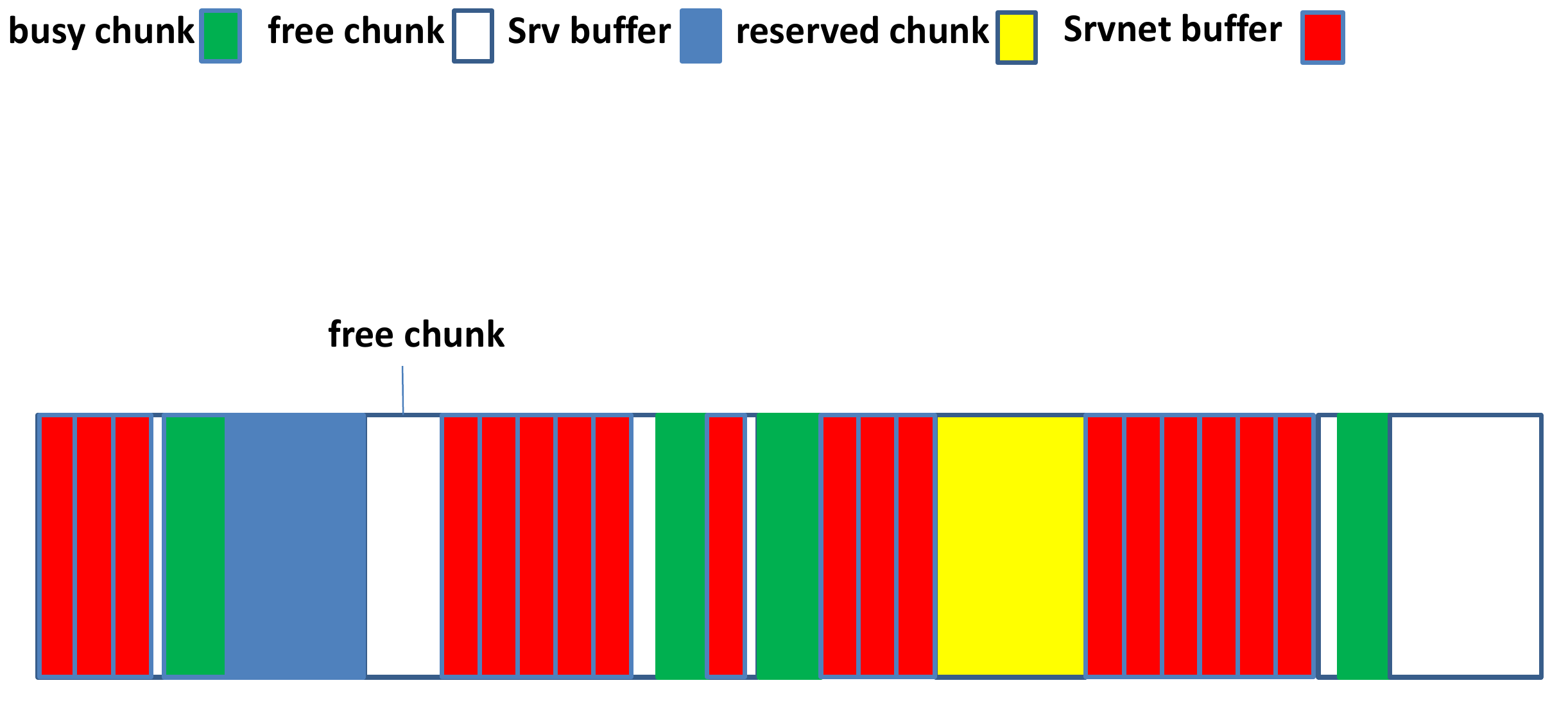}
    
      \caption{Server machine's updated buffer state}\label{buffer6}
      
   \end{figure}

Next, in connection No.17-22, extra Srvnet request packages are sent to the server. It is expected that one Srvnet.sys buffer allocated by these requests can be immediately after the \textit{result list buffer}, hence the overflow in the \textit{result list buffer} can overwrite the following Srvnet.sys buffer's header. Figure. \ref{buffer6} shows the server buffer status after receiving the extra Srvnet requests.

In connection No.16, the second reserved buffer is freed. And in connection No.1, the last segment of the original list is sent to the server, making the system start the conversion. 

\begin{figure}[ht]
      \includegraphics[scale=0.4]{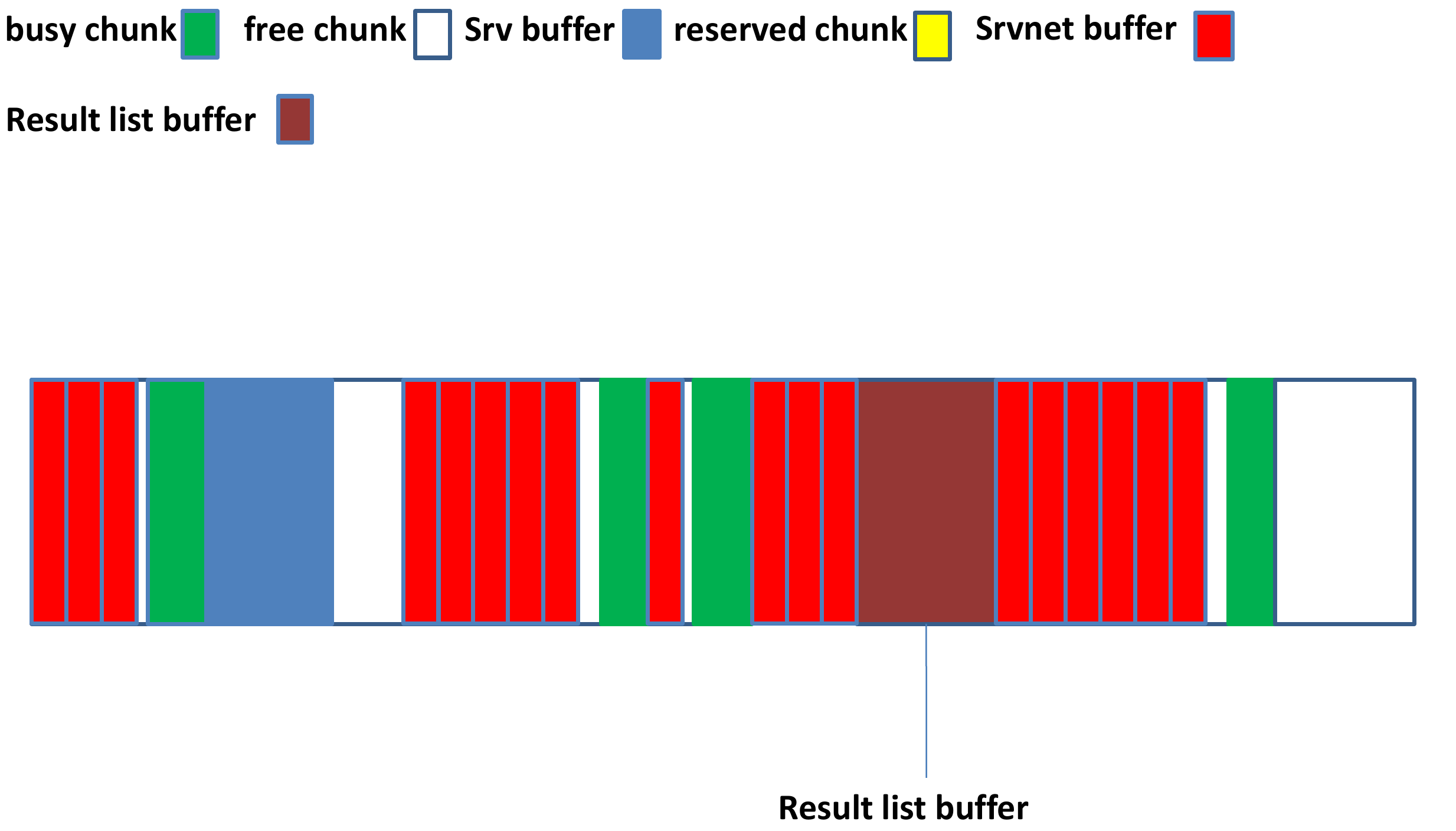}
    
      \caption{Server machine's updated buffer state}\label{buffer7}
      
   \end{figure}

To start the conversion process, the server tries to allocate a \textit{result list buffer} with size S1. Since the second reserved buffer, which has the size slightly greater than the result list, is just freed, Windows memory's last-in-first-out working fashion guarantees this buffer be allocated as the \textit{result list buffer}. During the conversion process, as introduced before, the data in the result buffer can overwrite the following Srvnet.sys buffer. The server buffer status changes to the one shown in Fig. \ref{buffer7}. 
 
Finally, the exploit sends the shellcode through each of the previously established Srvnet connections (i.e, connection No.3-15, and 17-22).
   
\subsection{Sending the shellcode}
The payload is a piece of executable code that is sent after the target machine is penetrated. Once the payload is executed, the attacker can leverage the vulnerability and do whatever he wants to do. In previous discussion, we have explained that by sending the grooming packages, multiple Srvnet connections are established and one of these corresponding Srvnet.sys buffes is expected to be overflown. According to the Check Point Reseach paper \cite{c5}, after the Srvnet connections are established, these connections wait for another data packages and upon receiving the data packet, it will map the data according to the parameter \textit{pMdl} contained in the Srvnet buffer's header. Also, upon closing the Srvnet connection, the function pointed by \textit{HandlerFunction} in the \textit{pSrvnetWskStruct} will be called. The Srvnet.sys buffer has a structure as shown in Algorithm \ref{srvnet}.

   \begin{algorithm}
   \caption{Srvnet.sys buffer structure}
   \label{srvnet}
   \begin{algorithmic}
   \STATE$Struct\ Srvnet\_header\{$
   \STATE$......$
   \STATE $MDL *pMdl1$
   \STATE $......$
   \STATE $Srvnet\_receive *pSrvnetWskStruct\{$
   \STATE $......$
   \STATE $PVOID HandleFunction$
   \STATE $\ \ \ \}$
   \STATE $......$
   \STATE $\} $
   \end{algorithmic}
   \end{algorithm}

In this section, we have introduced the \textit{shellcode sending process}. However, the detailed code analysis on the shellcode is not discussed in this paper as it is beyond the scope of this paper.

\section{Code analysis}

As described in Section 1, \textit{Wannacry} utilizes the famous exploit Eternalblue in its spreading process. In this section, we  provide a static analysis on Wannacry's exploit module to investigate how this exploit is utilized. The baseline of the analysis tools are shown below:
\begin{itemize}
  \item \textit{Wannacry} SHA256 hash: 24d004a104d4d54034dbcffc2a4b19a11\linebreak f39008a575aa614ea04703480b1022c
  \item Static analysis tool: IDA 6.8
\end{itemize}

\subsection{Wannacry}

Wannacry creates local network spreading threads and Internet spreading threads to propagate through the network. Both threads use the same exploit Eternalblue to infect other systems. 

In the local network spreading process, Wannacry creates a target IP address table and tries to attack the potential victims in the table exhaustively. In the Internet spreading process, Wannacry generates a random IP address and tries to attack the system sharing the same network segment. Like Eternalblue, the spreading process in Wannacry also consists of 3 essential steps: \textit{crafting original list}, \textit{buffer grooming} and \textit{sending the payload}. Table. \ref{table2} depicts the summery of Wannacry's package capture after we analyzed the network traffic during the infection. This table describes almost the same process as shown in Table. \ref{table1}, except for several differences. Even though the general process described in Table. \ref{table2} is similar to the process given in Table \ref{table1}, some of the packets are not introduced in Table. \ref{table1}, as they are unique in Wannacry.

\begin{table}[]
\caption{Summery of package capture of Wannacry}
\label{table2}
\begin{tabular}{|l|l|l|}
\hline
Step & Attempt                                                                                                 & Packages                                                                                                \\ \hline
1    & \begin{tabular}[c]{@{}l@{}}Detect the existence of MS17\_010\\ and DoublePulsar.\end{tabular}           & \begin{tabular}[c]{@{}l@{}}PeekNamedPipe Request and Trans2 Request\end{tabular}                      \\ \hline
2    & \begin{tabular}[c]{@{}l@{}}Send the original list except the \\ last segment.\end{tabular}               & \begin{tabular}[c]{@{}l@{}}A NT Trans Request and multiple Trans2 \\ Secondary Requests\end{tabular}   \\ \hline
3    & \begin{tabular}[c]{@{}l@{}}Ensure the package in last step\\ were sent successfully.\end{tabular}        & Echo Request                                                                                            \\ \hline
4    & Reserve the first buffer.                                                                               & \begin{tabular}[c]{@{}l@{}}Negotiate Protocol Request and Session Setup \\Andx Request\end{tabular} \\ \hline
5    & Reserve Srvnet.sys buffers.                                                                             & Multiple TCP packages                                                                                   \\ \hline
6    & Reserve the second buffer.                                                                              & \begin{tabular}[c]{@{}l@{}}Negotiate Protocol Request and Session Setup \\Andx Request\end{tabular} \\ \hline
7    & Free the first reserved buffer.                                                                         & A FIN TCP package                                                                                       \\ \hline
8    & Reserve extra Srvnet.sys buffers.                                                                       & Multiple TCP packages                                                                                   \\ \hline
9    & \begin{tabular}[c]{@{}l@{}}Ensure the packages sent in last\\ step were sent successfully.\end{tabular} & Echo Request                                                                                            \\ \hline
10   & Free the second reserved buffer.                                                                        & A FIN TCP package                                                                                       \\ \hline
11   & \begin{tabular}[c]{@{}l@{}}Send the last segment of the \\ original list.\end{tabular}                  & \begin{tabular}[c]{@{}l@{}}A Trans2 Secondary Request\end{tabular}                                   \\ \hline
12   & Send the shellcode.                                                                                     & Mutiple TCP packages                                                                                    \\ \hline
\end{tabular}
\end{table}

Through the static analysis of Wannacry by using IDA 6.8, we discovered the function beginning at offset 0x00401D80 crafts the fake original list and prepares the grooming packages, which is discussed in the \textit{crafting original list} and \textit{buffer grooming} steps. These fake original list, grooming packages, and the shellcode mentioned above are embedded into the ransomworm by the ransomworm author. During the runtime, they are extracted and pasted into a buffer chunk in the same order as listed in Table \ref{table1}. Then data in this buffer chunk will be sent later to the server, which spreads the ransomworm and executes the ransomworm on the server. All the data is in plain-text format and is barely different from the Metasploit's exploit. We will discuss the particulars in the following paragraphs.

\begin{figure}[ht]
      \includegraphics[scale=0.75]{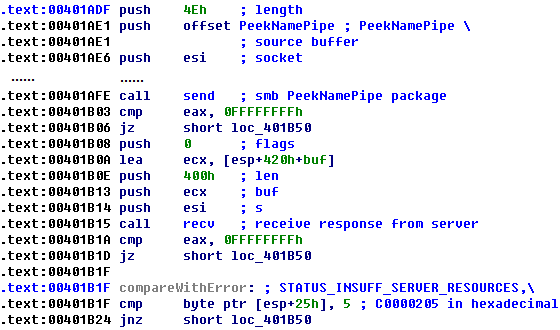}
    
      \caption{PeekNamePipe package}\label{peekWN}
      
   \end{figure}

\begin{figure}[ht]
      \includegraphics[scale=0.75]{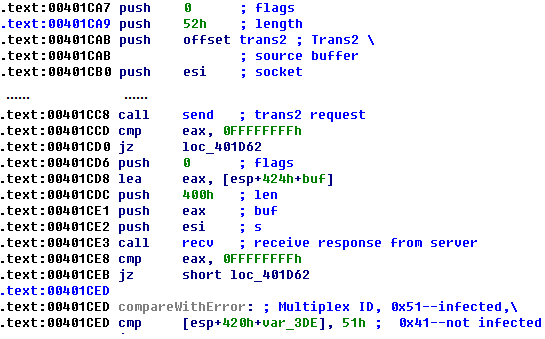}
    
      \caption{Trans2 package}\label{TransWN}
      
   \end{figure}

Prior to preparing the grooming packages and the shellcode, Wanancry sends a PeekNamePipe package and a Trans2 Request package to detect the existence of MS17\_010 and backdoor Doublepulsar respectively \cite{c14} as in Fig. \ref{peekWN} and Fig. \ref{TransWN}. As step 1 listed in Table. \ref{table2}, the PeekNamePipe package data is embedded into the ransomworm, as depicted in Fig. \ref{peekWN}. It is used when the ransomworm needs to send it (by instruction \textit{call send} at offset 00401AFE). After sending this package, the ransomworm waits for the response from the server by calling the \textit{recv} function at offset 00401B15. If the data in the response package equal to \textit{STATUS\_INSUFF\_SERVER\_RESOURCES} (\textit{0xC0000205} in hexadecimal), that denotes the MS17\_010 vulnerability resides on the server. 
As shown in Fig. \ref{TransWN}, the Trans2 package data is also embedded into the ransomworm. The ransomworm waits for the response from the server by the instruction \textit{call recv}. If the \textit{Multiplex ID} field in the response package equals to \textit{0x51}, that denotes the server is infected with Doublepulsar, whereas if the field equals to \textit{0x41}, that denotes the server is not infected.    

\begin{figure}[ht]
      \includegraphics[scale=0.75]{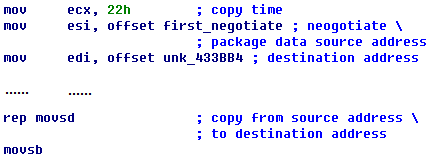}
    
      \caption{Negotiate Protocol package}\label{1stneo}
      
   \end{figure}

\begin{figure}[ht]
      \includegraphics[scale=0.75]{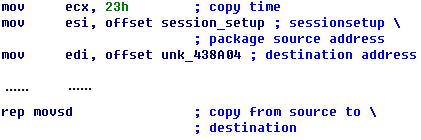}
    
      \caption{Session Setup AndX package}\label{1stses}
      
   \end{figure}

\begin{figure}[ht]
      \includegraphics[scale=0.75]{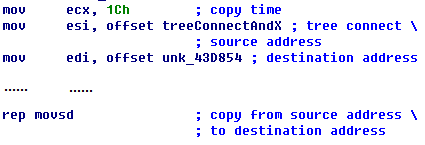}
    
      \caption{Tree Connect AndX package}\label{1sttree}
      
   \end{figure}

To establish the connection to the server (victim) machine, the first Negotiate Protocol package is crafted as shown in Fig. \ref{1stneo}. The Session Setup AndX and Tree Connect AndX Request packages are crafted in the similar way as depicted in Fig. \ref{1stses} and Fig. \ref{1sttree}.
   
\begin{figure}[ht]
      \includegraphics[scale=0.65]{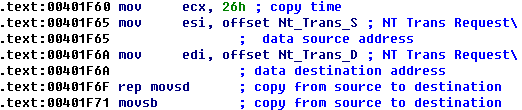}
    
      \caption{Nt Trans Request}\label{nttrans}
      
   \end{figure}

\begin{figure}[ht]
      \includegraphics[scale=0.75]{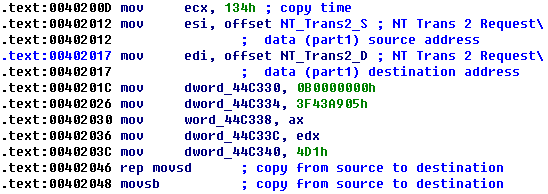}
    
      \caption{Part of Nt Trans2 Request}\label{nttrans2}
      
   \end{figure}

\begin{figure}[ht]
      \includegraphics[scale=0.75]{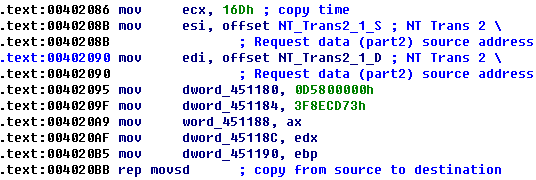}
    
      \caption{Part of Nt Trans2 Request}\label{nttrans2_1}
      
   \end{figure}

Next, as in step 2 of Table. \ref{table2}, an NT Trans Request package, and multiple Trans2 Secondary Request packages containing the crafted Os2Fea list without the last segment are prepared as in Fig. \ref{nttrans}, Fig. \ref{nttrans2}, and Fig. \ref{nttrans2_1}. 

\begin{figure}[ht]
      \includegraphics[scale=0.75]{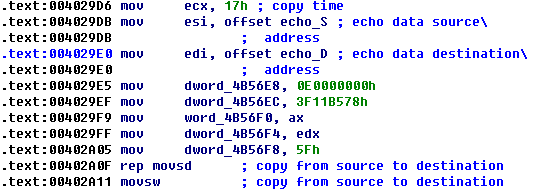}
    
      \caption{Echo package to check original list well received} \label{echocode}
      
   \end{figure}

In step 3 of Table. \ref{table2}, to ensure the original list is received completely, an echo package is prepared as shown in Fig. \ref{echocode}. 

\begin{figure}[ht]
      \includegraphics[scale=0.65]{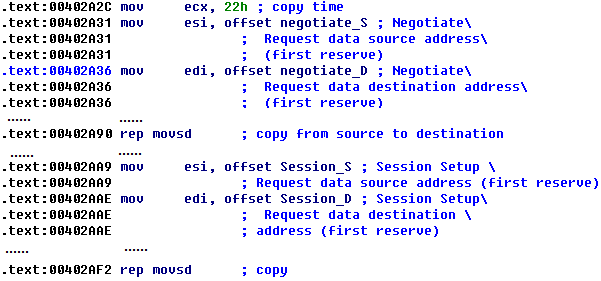}
    
      \caption{Grooming package reserves a buffer chunk}\label{1streservecode}
      
   \end{figure}

In step 4 of Table. \ref{table2}, a package which reserves the first buffer chunk on the target is prepared. The corresponding Negotiate and Session Setup Request packages are shown in Fig. \ref{1streservecode}. 

\begin{figure}[ht]
      \includegraphics[scale=0.75]{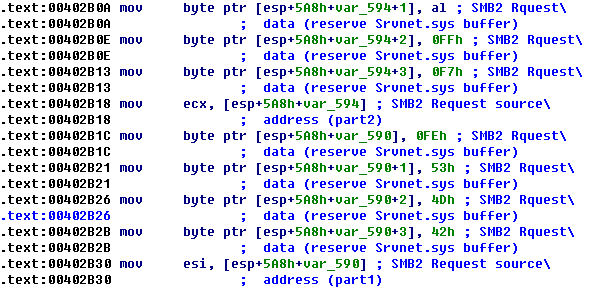}
    
      \caption{Crafting grooming package which reserves a buffer chunk}\label{srvnetcode0}
      
   \end{figure}

As in step 5 of Table. \ref{table2}, to continue the buffer grooming process, multiple Srvnet connection requests should be sent to reserve Srvnet.sys buffer chunks on the target system. Figure. \ref{srvnetcode0} shows the crafting of each Srvnet package. 

   \begin{figure}[ht]
      \includegraphics[scale=0.75]{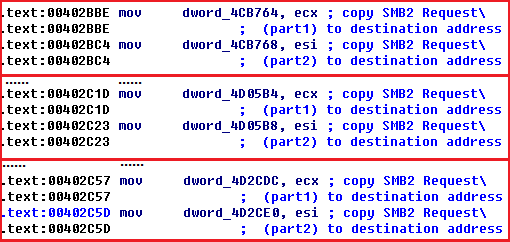}
    
      \caption{Copying crafted grooming package to destination address}\label{srvnetcode}
      
   \end{figure}

\begin{figure}[ht]
      \includegraphics[scale=0.75]{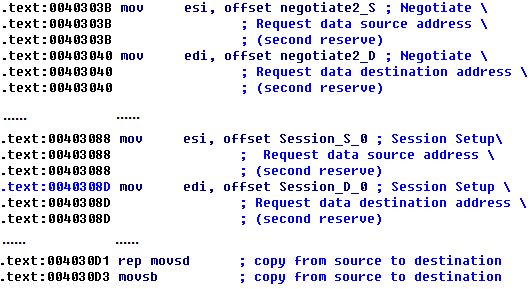}
    
      \caption{Second package to reserve a buffer chunk}\label{2ndreservecode}
      
   \end{figure}

As in step 6 of Table. \ref{table2}, the second reserving package shall be sent to the server. The crafting process is shown in Fig. \ref{2ndreservecode}, including Negotiate Protocol Request and Session Setup AndX Request. In step 7, the first reserving buffer allocated previously shall be freed. 

\begin{figure}[ht]
      \includegraphics[scale=0.75]{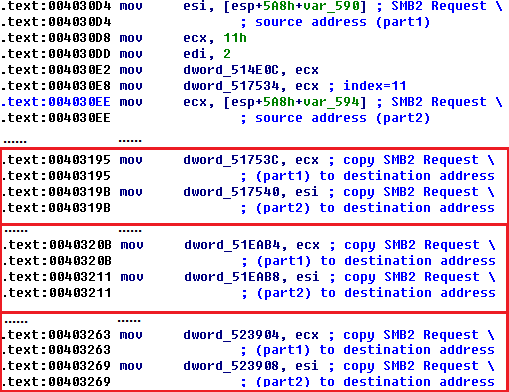}
    
      \caption{Crafting extra SMB2 requests which reserve Srvnet.sys buffer}\label{extrasrvnetcode}
      
   \end{figure}

\begin{figure}[ht]
      \includegraphics[scale=0.75]{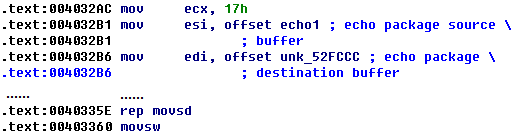}
    
      \caption{Crafting Echo request}\label{wn_echo1}
      
   \end{figure}

In step 8, extra 5 Srvnet connection requests are crafted as shown in Fig. \ref{extrasrvnetcode} and will be sent to the target machine to reserve Srvnet.sys buffers. It is expected that one Srvnet.sys buffer allocated in this step is immediately after the second reserved buffer, which will be replaced with the \textit{result list buffer} later. In step 9, an Echo package is crafted as shown in Fig. \ref{wn_echo1}.

\begin{figure}[ht]
      \includegraphics[width=1\linewidth]{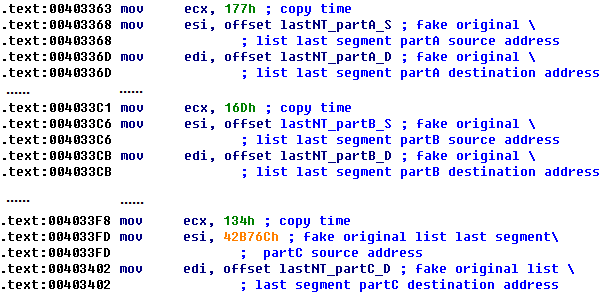}
    
      \caption{Last segment of original list}\label{lastsegcode}
      
   \end{figure}

In step 10, the second reserved buffer shall be freed and in step 11, the last segment of the crafted original list shall be sent to the target as shown in Fig. \ref{lastsegcode}. Once this last segment is received, the target's conversion process (converting the original list to the result list) begins. 

\begin{figure}[ht]\centering 
      \includegraphics[width=1.05\linewidth]{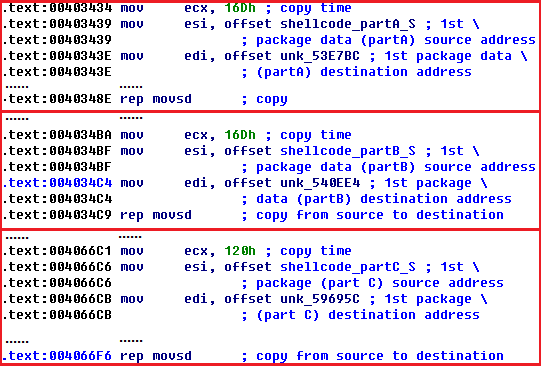}
    
      \caption{Copying part A, B and C of the first shellcode package}\label{1stpackage0}
      
   \end{figure}

In step 12, multiple packages that contain the same shellcode are crafted as Figs. \ref{1stpackage0}. Later they will be sent through the Srvnet connections established earlier. 

\begin{figure}[ht]\centering 
      \includegraphics[width=\linewidth]{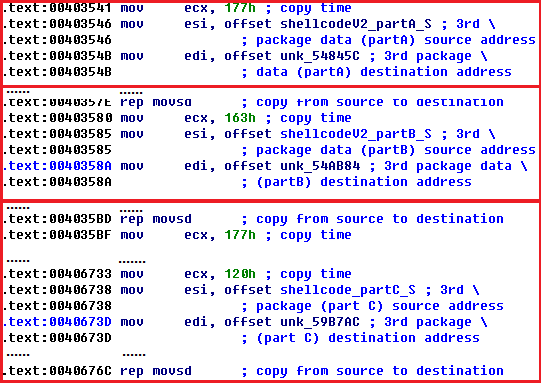}
    
      \caption{Preparing part A, B and C of the third shellcode package}\label{3rdpackage0}
      
   \end{figure}

\begin{figure}[ht]\centering 
     \includegraphics[width=1.1\linewidth]{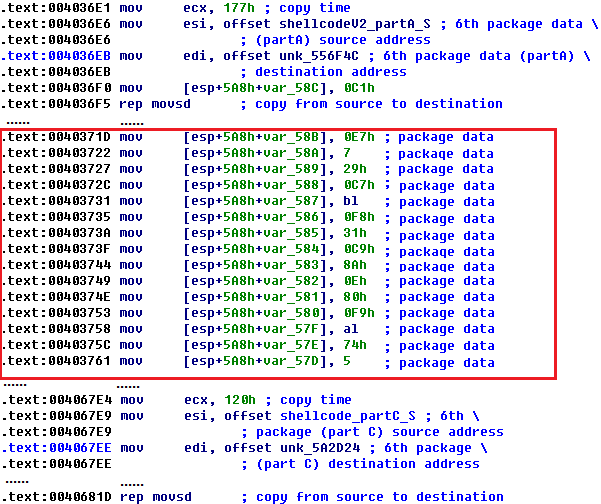}
    
      \caption{Part A and B of the sixth shellcode package is prepared}\label{6thpackage0}
      
   \end{figure}

~\\
~\\
~\\
~\\
~\\
~\\
~\\
~\\
~\\
~\\
~\\
~\\
~\\
~\\
\section{conclusion}
The wide application of exploit Eternalblue is a meaningful security incident. The massive infection based on Eternalblue spurs everyone to raise the awareness of patching computers to current status. This paper introduced the underlying mechanism of exploit Eternalblue, as well as the reverse engineering result of Eternalblue. The code of Eternalblue applied in Wannacry is compared with the original exploit based on the reverse engineering results. The analysis reveals that the exploit Eternalblue is slightly modified when applied in Wannacry. Our work gathered much-known knowledge of Eternalblue to provide readers with a clear picture of this exploit. We have concluded the similarity and the difference of Eternalblue's code in Wannacry. We have also analyzed Notpetya and found the exploit in Notpetya is encrypted and only decrypted while the shellocode is executed. After decrypting it, the Notepetya exploit is almost identical to the original Eternalblue exploit. Due to the length constrains of the paper, the analysis details are not included here. It is  possible to extend our work to the code analysis for ransomworm detection. 

\bibliographystyle{splncs04}
\bibliography{v1_main.bbl}
\end{document}